\documentclass[10pt,leqno]{article}
\usepackage{graphicx}
\usepackage{indentfirst,csquotes}

\usepackage{amssymb,amsthm,amsmath}
\usepackage{xcolor,paralist,hyperref,titlesec,fancyhdr,etoolbox}

\hypersetup{ colorlinks=true, linkcolor=black, filecolor=black, urlcolor=black }

\usepackage{lipsum}

\usepackage{authblk}
\usepackage{titlesec}

\usepackage{chemformula} % Formula subscripts using \ch{}

\usepackage{textgreek}

\usepackage[separate-uncertainty=true]{siunitx}
\DeclareSIUnit\Molar{\textsc{m}}
\DeclareSIUnit\eq{eq.}

\usepackage{floatrow}
\usepackage[sorting=none]{biblatex}
\begin{document}
\title{Hydrothermal Synthesis of Ultra-high Aspect Ratio \textbeta-NaYF Disks via Methyliminodiacetic Acid (MIDA)} %%%%%%%%%%%%

\author[1]{Lars Forberger}
\author[2]{Jacob T. Baillie}
\author[1]{Zhaojie Feng}
\author[1]{Rachel E. Gariepy}
\author[1]{Sankhya Hirani}
\author[2]{Daniel R. Gamelin}
\author[1, 3]{Shuai Zhang}
\author[2]{Werner Kaminsky}
\author[*1, 4]{Peter J. Pauzauskie}
\affil[1]{Department of Materials Science \& Engineering, University of Washington, Seattle, WA 98195, USA}
\affil[2]{Department of Chemistry, University of Washington, Seattle, WA 98195, USA}
\affil[3]{Physical Sciences Division, Pacific Northwest National Laboratory, Richland, WA 99354, USA}
\affil[4]{Physical and Computational Sciences Directorate, Pacific Northwest National Laboratory, Richland, WA 99354, USA}
\affil[*]{Email Address: peterpz@uw.edu}

\maketitle

\begin{abstract}
The hexagonal \textbeta-phase of sodium yttrium fluoride (\ch{NaYF}) is a leading host material for
lanthanide upconversion and anti-Stokes fluorescence laser refrigeration based on its low phonon energies and high upconversion efficiency. Recently experiments have been proposed to use this material as an optically-levitated sensor of high-frequency gravitational waves. In order to maximize signal-to-noise in this experiment, the \ch{NaYF} sensor must have both a two-dimensional, disk-like morphology 
and also a large mass. Here we report a novel hydrothermal process based on the chelation ligand methylimidodiacetic acid (\ch{MIDA}) to realize 
hexagonal \ch{\textbeta-NaYF} prisms with corner-to-corner diameters up to \SI{44}{\micro m} while keeping the height around \SI{1}{\micro m}.
The surface quality is comparable to particles synthesized with \ch{EDTA} based on atomic force microscopy (AFM) measurements. Unlike particles synthesized with \ch{EDTA} the \ch{\textbeta-NaYF} particles show no lensing based on curvature of the hexagonal basal plane. Single crystal X-ray diffraction data were refined to the P$\bar{6}$2c (\#190) space group which to the best of our knowledge has not been reported in the literature. One of six \SI{44}{\micro m} \ch{\textbeta-NaYF} disks doped with \SI{10}{\percent} ytterbium showed laser refrigeration of \SI{-4.9(10)}{K} suggesting future applications in both levitated optomechanics and microoptics.
\end{abstract} %%%%%%%%%

\bigskip

\section{Introduction}
Upconversion materials are widely studied for potential applications such as physical sensors, biosensors, artificial light sources, displays, anti-counterfeiting agents and solar cell enhancers \cite{Chen2020, Malhotra2023, Zhang2022, Zhu2019, Zheng2022, Khare2020}. The materials developed for most of these applications are on the nanoscale and doped with combinations of lanthanides to achieve visible output following near-infrared (NIR) excitation based on multiphoton upconversion. Microparticles are less suitable for some of the aforementioned applications but are more relevant in microoptics, e.g. as microlaser cavities \cite{Wang2017, FernandezBravo2018, Liu2020}. Using ytterbium (\ch{Yb^{3+}}) as a dopant furthermore adds the possibility of anti-Stokes fluorescence laser refrigeration offering additional hypothetical uses like temperature controlled micromirrors, microlenses, waveguides, radiation balanced lasers, and payload cooling substrates \cite{Hehlen2018, Meng2018}. Levitated-sensor applications benefit from laser refrigeration as well due to the avoidance of the negative effects associated with heating, like thermal noise, motional instability, and sample degradation. Examples include high-precision accelerometers and the detection of high frequency gravitational waves using high-mass sensors optically levitated in two arms of a Michelson interferometer \cite{Aggarwal2022}. Many of these systems require specific particle morphologies, sizes, and masses in order to trap the sensors, reach sensitivity goals or generate the desired motional dynamics\cite{Winstone2022}. Here, we focus on the design of high-mass hexagonal disks that can be trapped in the antinode of a dual-beam standing wave trap, limiting the particle thickness to roughly \SI{1}{\micro m}. To increase the mass of the particles, we aim to maximize the corner-to-corner diameter and accordingly diameter-to-thickness aspect ratio (AR), while maintaining the laser refrigeration capabilities made possible through ytterbium (\ch{Yb^{3+}}) doping. The cooling and disk-like morphology enhance the readout quality by reducing the thermal noise and maximizing light scattering in the forward direction where detection occurs \cite{Aggarwal2022, Winstone2022}.

\begin{figure}[b!]
	\includegraphics[width=0.9\linewidth]{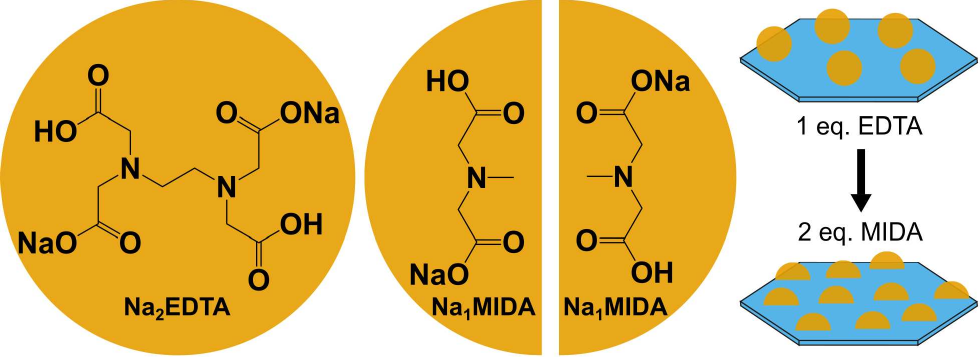}
    \centering
	\caption{
		Chemical structures of \ch{EDTA} and \ch{MIDA}. Illustration of the hypothetical surface 
		coverage increase on a \ch{\textbeta-NaYF} nucleus by substituting \ch{EDTA} with two 
		equivalents of \ch{MIDA}.
		}
	\label{figure:MIDA_Concept}
\end{figure}

Since its original discovery \cite{Thoma1963, Menyuk1972}, the hexagonal \textbeta-phase of sodium yttrium fluoride (\ch{\textbeta-NaYF}) has been demonstrated to be one of the most efficient upconversion host materials and has been studied extensively \cite{Haase2011}. Single crystals of \ch{\textbeta-NaYF} can be synthesized via multiple different routes, including melt growth, thermal decomposition, coprecipitation and solvo- or hydrothermal methods  \cite{Knowles1989, Aebischer2006, Zhu2019}. Melt growth approaches to obtain macroscopic samples have failed in the past \cite{Knowles1989, Aebischer2006}, resulting in upper size limits of single
crystalline particles in the micron range. These are most commonly obtained using hydrothermal synthesis, where the size and morphology are controlled by the addition of organic molecules. These often have a twofold functionality, acting as chelating agents as well as surface adsorbents that alter the nucleation and 
growth thermodynamics and kinetics. The most commonly used organic additives are sodium citrate 
and ethylenediaminetetraacetic acid (\ch{EDTA}). 
We are unaware of synthesis optimizations towards hexagonal disks and the typical diameter obtained is to the best of our knowledge below \SI{5}{\um} \cite{Bogachev2022, Chen2015a, Chen2012, Fu2017, Hao2012, Li2007, OrtizRivero2021, Sahoo2022a, Som2021, Wang2015a, Wang2013, Yang2012a}.
Felsted et al. showed that EDTA can be used to tune the morphology of 
\ch{\textbeta-NaYF} fluently from hexagonal plates to rods by changing the ratio of rare-earth (RE) to fluoride ions. The largest 
hexagonal disk obtained had a diameter of \SI{5}{\micro m} and $\sim$\SI{500}{nm} heights, resulting in an aspect ratio of \num{10.5} \cite{Felsted2022}.
 
The fundamental idea of this work is to increase the surface coverage of ligands during growth by substituting one molecule of \ch{EDTA} with two molecules of methyliminodiacetic acid (\ch{MIDA}). \ch{MIDA} is 
effectively half an \ch{EDTA} molecule, and two equivalents should in principle be able to yield a 
comparable control over the particle morphology. While the RE-\ch{MIDA} complex will be less stable than the 
RE-\ch{EDTA} equivalent for entropic reasons \cite{Smith1987}, the surface coverage on a flat surface should be 
increased, as illustrated in \textbf{Figure \ref{figure:MIDA_Concept}}. Furthermore, \ch{MIDA} has a higher 
solubility than \ch{EDTA} allowing a wider range of pH values to be explored within the available thermodynamic parameter space. Both \ch{EDTA} and \ch{MIDA} exist in different ionic species depending on the pH of the solution due to the acidic carboxylic acid and basic amine groups as shown for \ch{MIDA} in \textbf{Figure \ref{figure:SynthesisOptimizationSEM}}. 
Here, the pH-dependent morphology control of \ch{\textbeta-NaYF} during the \ch{MIDA}-assisted hydrothermal synthesis is presented. At pH values below the inflection point, rods are obtained. Increasing the amount of \ch{NaOH} to \SI{1}{\eq} yields hexagonal disks with corner-to-corner width up to \SI{44}{\micro m}. Beyond the inflection point, semicircular disks were observed before hexagonal prisms were obtained in the high-pH regime.
Single crystal X-ray diffraction of individual \ch{Na_{1.00}MIDA} \ch{\textbeta-NaYF} disk indicate they have a P$\bar{6}$2c space group that, to the best of our knowledge, has not yet been reported. Further, one of six measured particles showed laser refrigeration of \SI{-4.9(10)}{K} opening avenues towards future solid-state laser refrigeration of these microdiscs within both optically-levitated gravitational wave interferometers and integrated optoelectronic devices.

\section{Results and Discussion}
\begin{figure}[ht!]
	\includegraphics[width=0.9\linewidth]{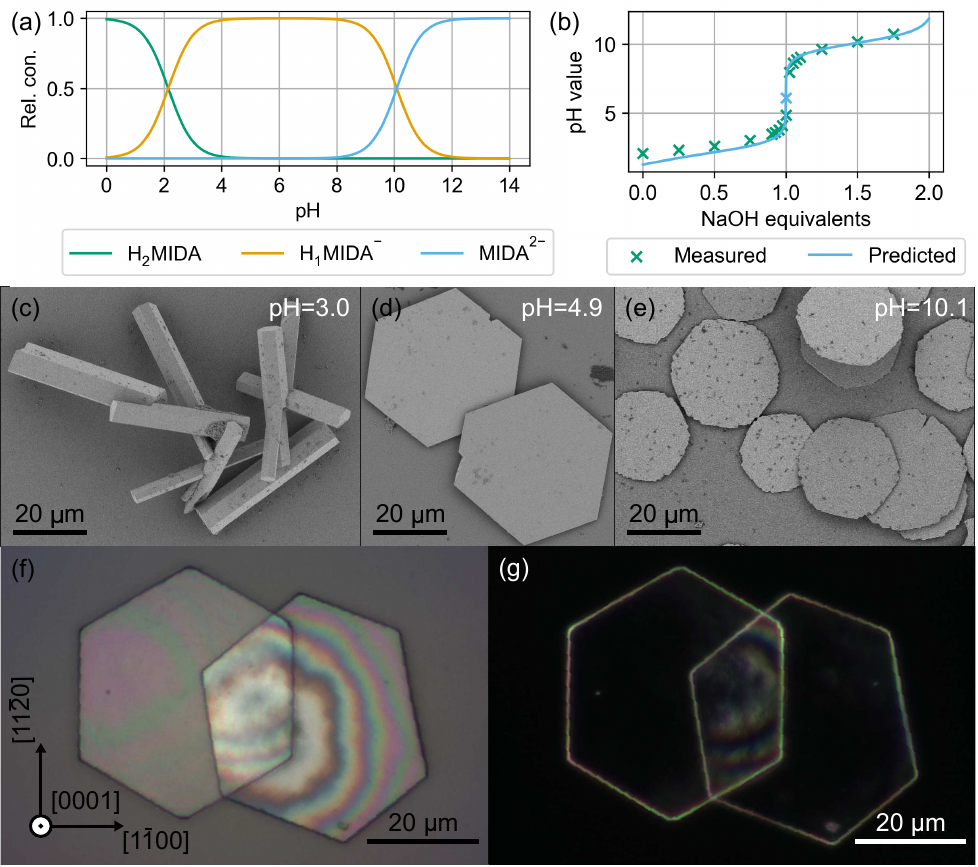}
    \centering
	\caption{
		Calculated relative abundances of the different ionic \ch{MIDA} species depending on pH (a). Predicted and measured pH value for a \ch{MIDA} solution after the addition of \ch{NaOH}. The inflection point, marked as blue cross, is calculated to be at a pH value of \num{6.1}. The p$K_a$ values of \ch{MIDA} are \num{2.146} and \num{10.088} (b) \cite{Ockerbloom1956}.
        Hydrothermal synthesis results using MIDA in different protonation states by the addition 
		of \SI{0.75}{\eq} (c), \SI{1.00}{\eq} (d) and  \SI{1.50}{\eq} (e) of sodium hydroxide.
        Optical bright- (f) and darkfield (g) images of the \ch{\textbeta-NaYF} disks obtained with \ch{Na_{1.00}MIDA}. Crystallographic directions for the left particle are given in the optical brightfield image. 
		}
	\label{figure:SynthesisOptimizationSEM}
\end{figure}

The influence of MIDA as an organic additive on the size and morphology of hydrothermally grown \textbeta-NaYF was investigated and compared to \ch{EDTA}. During the optimization, aiming for high aspect ratio disks with heights below \SI{1}{\um}, different protonation states of \ch{MIDA} were studied by dissolving the ligand with different amounts of sodium hydroxide (\ch{NaOH}) ranging from \SIrange{0}{2}{\eq} Since the \ch{MIDA} titration curve has an inflection point at \SI{1}{\eq} \ch{NaOH} with a predicted pH value of \num{6.12}, as shown in Figure \ref{figure:SynthesisOptimizationSEM}, two syntheses with pH values just above and below the inflection point were run.
The amount of \ch{NaOH} used heavily affects the phase distribution, size and morphology of the product as shown in \textbf{Table \ref{table:NaOHOptimizationResults}}. Powder X-ray diffraction (XRD) was used to determine the synthesized phase(s) and scanning electron microscopy (SEM) was used to assess the particle morphology and size.
While \ch{\textbeta-NaYF} is obtained 
under all conditions, the addition of low amounts of \ch{NaOH} ($\leq$ \SI{0.50}{\eq}) led to the 
formation of yttrium trifluoride (\ch{YF3}) as byproduct. The syntheses using \SIrange{0.75}{1.50}{\eq} resulted in mixed \textalpha- and \textbeta-phases with the exception of \ch{Na_{1.00}MIDA} above the pH inflection point. The \ch{\textalpha-NaYF} particles were sub \SI{100}{nm} spheres, very comparable to those obtained in hydrothermal synthesis at lower temperatures, typically \SI{90}{\degreeCelsius} using \ch{EDTA} and otherwise similar conditions \cite{LuntzMartin2021}. \SI{2.00}{\eq} \ch{NaOH} was the only amount consistently yielding phase-pure \ch{\textbeta-NaYF}.
Due to the large mass difference between the \ch{\textbeta-NaYF} microparticles and \ch{\textalpha-NaYF} nanoparticles a simple separation by supernatant removal from \ch{EtOH} dispersions was used to obtain the hexagonal disks prior to any further characterization (see \textbf{Figure S1}).
From the lowest to highest amount of added \ch{NaOH} the 
\ch{\textbeta-NaYF} particle morphology changed from rods to hexagonal disks to semicircular disks and eventually back to hexagonal disks (see \textbf{Figure} \ref{figure:SynthesisOptimizationSEM} and \textbf{S2}). It is noteworthy that the organic redsidue appearing black in the SEM images is from the EtOH used for dropcasting and not a synthesis residue as illustrated in \textbf{Figure S3}. The prisms obtained with \SI{2.00}{\eq} of \ch{NaOH} 
appear to be several fused hexagonal disks (Figure S2 (d)) while the particles obtained at \SI{1.00}{\eq} show very clear hexagonal faceting without any obvious defects visible in SEM or optical microscope images (Figure \ref{figure:SynthesisOptimizationSEM} ((d), (f) and (f)). The optical microscopy images further illustrate the transparency of \ch{\textbeta-NaYF} in the visible spectrum and point towards applications as microoptics. From \SIrange{0.00}{0.75}{\eq} \ch{NaOH} the hexagonal rods were obtained. The width increased from \SI{3.4}{\micro m} for the first two amounts to 
\SI{8.7}{\micro m}, and the length from \SI{15.6}{\micro m} over \SI{16.8}{\micro m} to \SI{42.1}{\micro 
m} (Table \ref{table:NaOHOptimizationResults}, Figure \ref{figure:SynthesisOptimizationSEM} (c) and Figure S2 (a), (b)) resulting in a width-to-height aspect ratio (AR) of roughly \num{0.2} for all rod-like morphologies. 
The hexagonal disks obtained with \SI{1.00}{\eq} of \ch{NaOH} have an aspect ratio of \num{44} 
(\SI{44.0}{\micro m} $\times$ \SI{1.0}{\micro m}) or \num{21.1} (\SI{38.0}{\micro m} $\times$ \SI{1.8}{\micro m}), while the disks obtained with \SI{2.00}{\eq} \ch{NaOH} result in \SI{7.5}{\micro m} by \SI{1.3}{\micro m} particles (AR: \num{5.7}). 
The semicircular disks obtained with \SI{1.25}{\eq} and \SI{1.50}{\eq} \ch{NaOH} yielded particles with aspect ratios of \num{35.8} (\SI{14.3}{\micro m} $\times$ \SI{0.4}{\micro m}) and \num{21.0} (\SI{25.2}{\micro m} 
$\times$ \SI{1.2}{\micro m}) as shown in Figure \ref{figure:SynthesisOptimizationSEM} (e) and Figure S2 (c). 

The main goal for the synthesis optimization was the growth of high aspect ratio hexagonal \ch{\textbeta-NaYF} disks with heights below \SI{1}{\um} as levitated sensors in dual beam optical traps. The particles obtained with \ch{MIDA} and the addition of \SI{1.00}{\eq} of \ch{NaOH} resulting in a pH just below the inflection point are \SI{44}{\um} wide and \SI{1}{\um} high, resulting in an aspect ratio of 44, which is an increase in width of over eight and in aspect ratio of over four times greater compared to previously reported syntheses using \ch{EDTA}, citrate or other organic additives.
The crystallographic orientation of the particles, indicated in Figure \ref{figure:SynthesisOptimizationSEM} (f), was investigated using electron backscatter diffraction (EBSD), the results are summarized in \textbf{Figure S4}.
The minimum growth velocity over \SI{72}{h} is calculated to be \SIrange{260}{310}{nm.h^{-1}} along the \{11$\bar{2}$0\} directions and \SIrange{225}{270}{nm.h^{-1}} for the \{1$\bar{1}$00\} family of directions. The growth velocity towards the basal plane \{0001\} family of directions on the other hand is only \SIrange{5}{15}{nm.h^{-1}}.
Interestingly, the best results for both \ch{MIDA} and \ch{EDTA} are obtained when half of the carboxylic acid groups are deprotonated, i.e. \SI{1.00}{\eq} \ch{NaOH} per \ch{MIDA} molecule or \SI{2.00}{\eq} \ch{NaOH} per \ch{EDTA} molecule. 

\begin{table}[bt]
	\caption{Experimental conditions and results of the MIDA assisted hydrothermal 
	\ch{\textbeta-NaYF} 
	synthesis.}
	\label{table:NaOHOptimizationResults}
    \centering
	\begin{tabular}{cccccc}
		\hline
		NaOH [\si{\eq}] & pH & Phase & Morphology & Size [\si{\um}] & Aspect Ratio \\
		\hline
		\num{0.00} & 2.1 & \ch{YF3}, \textbeta & Rods & 
		\num{3.1 \pm 1.4} $\times$ \num{15.6 \pm 5.7} & 0.2 \\
		\num{0.50} & 2.6 & \ch{YF3}, \textbeta & Rods &  
		\num{3.1 \pm 0.6} $\times$ \num{16.8 \pm 4.2} & 0.2 \\
		\num{0.75} & \num{3.0} & \textalpha, \textbeta & Rods & 
		\num{8.7 \pm 1.9} $\times$ \num{42.1 \pm 3.5} & 0.2 \\
		\num{1.00} & \num{4.9} & \textalpha, \textbeta & Hexagonal disks & \num{44.0 \pm 1.1} $\times$ \num{1.0 \pm 0.1} & 44.0 \\
        \num{1.00} & \num{7.7} & \textbeta & Hexagonal disks & 
		\num{38.0 \pm 5.2} $\times$ \num{1.8 \pm 0.3} & 21.1 \\
		\num{1.25} & \num{9.6} & \textalpha, \textbeta & Circular disks &
		\num{14.3 \pm 6.8} $\times$  \num{0.4 \pm 0.2}  & 35.8 \\
		\num{1.50} & \num{10.1} & \textalpha, \textbeta & Circular disks & 
		\num{25.2 \pm 2.4} $\times$ \num{1.2 \pm 0.1} & 21.0  \\
		\num{2.00} & \num{11.7} & \textbeta & Hexagonal disks & 
		\num{7.4 \pm 0.7} $\times$ \num{1.3 \pm 0.2} & 5.7 \\
		\hline
	\end{tabular}
\end{table}

As mentioned in the introduction, \ch{MIDA} offers a bigger optimization parameter space than \ch{EDTA} due to being more soluble in water. This allows for the growth of \ch{\textbeta-NaYF} microparticles with different morphologies, from rods over disks to semicircular geometries. Unfortunately, the adjustability regions appear to be rather small, complicating the optimization for any desired particle morphologies or dimensions. This is especially true for the \SI{1}{\eq} synthesis, as illustrated by the different results obtained just above and below the pH inflection point of \ch{MIDA}. Furthermore, the use of \ch{MIDA} resulted in phase mixtures for most of the experiments carried out here, which is not commonly observed when using \ch{EDTA}. Lastly, the \ch{MIDA} system is much more dependent on initial conditions than the \ch{EDTA} counterpart. Using old Teflon liners (used once for another \ch{MIDA} trial) led to observable changes in size, morphology and aspect ratio. A \SI{1.00}{\eq} \ch{MIDA} synthesis with pH of \num{4.1} resulted in inhomogeneous hexagonal disks with aspect ratios of \num{24} and sizes of (\SI{30.9}{\micro m} $\times$ \SI{1.3}{\micro m}) compared to (\SI{44.0}{\micro m} $\times$ \SI{1.0}{\micro m}, AR: 44). In typical \ch{EDTA} syntheses, the morphologies and aspect ratios are maintained and only the overall particle size is affected. These behaviors could be caused by altered heterogeneous nucleation rates as a result of an increase in the surface area and roughness of the Teflon liners. 

\begin{figure}[t]
	\includegraphics[width=0.9\linewidth]{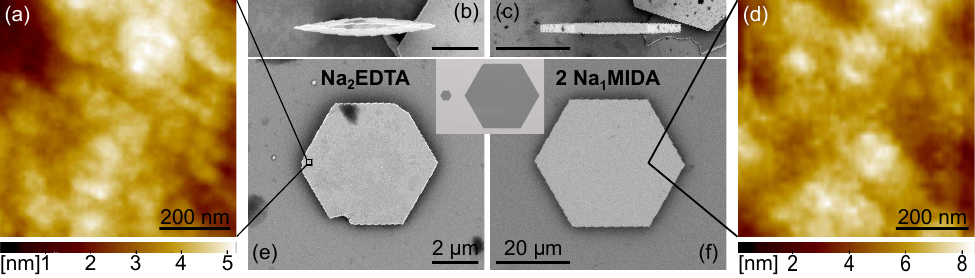}
    \centering
	\caption{Size, morphology and surface comparison of \ch{Na_{2.00}EDTA} \ch{\textbeta-NaYF} ((a), (b), (e)) and \ch{Na_{1.00}MIDA} \ch{\textbeta-NaYF} ((c), (d), (f)). AFM surface characterization ((a), (d)), SEM side ((b), (c)) and top ((e), (f)) view, the inset illustrates the particle dimensions to scale.}
	\label{figure:Radius of Curvature}
\end{figure}

Hexagonal \ch{\textbeta-NaYF} disks grown using \ch{EDTA} have the tendency to show curvature across the basal plane of the hexagonal disk, also known as lensing. It has been hypothesized that the curvature is due to terrace growth \cite{Felsted2022}.
Basal plane curvature is significantly reduced in hexagonal disks synthesized using \ch{MIDA} as shown in \textbf{Figure \ref{figure:Radius of Curvature}}. While the \ch{MIDA} disks are virtually flat (radius of curvature approaching infinity) the synthesized \ch{EDTA} microparticles show increased heights toward the center of the disks corresponding to a radius of curvature of roughly \SI{14}{\um}. Both morphologies have potential applications as laser refrigerated microoptics, the \ch{EDTA} particles are biconvex lenses with a calculated focal distance of roughly \SI{14.4}{\micro m}. In contrast the flat surface of the \ch{MIDA} disks could enable planar waveguiding of four different optical modes at a wavelength of \SI{1020}{nm}. 
Atomic force microscopy (AFM) was used to compare the surfaces resulting from the different organic additives. For three different positions on two \ch{Na_{2.00}EDTA} \ch{\textbeta-NaYF} disks a roughness average 
($R_a$) of \SI{4.2 \pm 1.8}{nm} and root-mean-square roughness ($R_q$) of \SI{4.3 \pm 1.8}{\nm} 
were obtained. A similar analysis of \ch{Na_{1.00}MIDA} disks yielded an $R_a$ of \SI{4.7 \pm 1.2}{nm}  and 
an $R_q$ of \SI{4.9 \pm 1.1}{nm} showing a comparable surface roughness when using \ch{MIDA} instead of \ch{EDTA}, as shown in Figure \ref{figure:Radius of Curvature} (a) and (d). 

\begin{figure}[t]
	\includegraphics[width=0.9\linewidth]{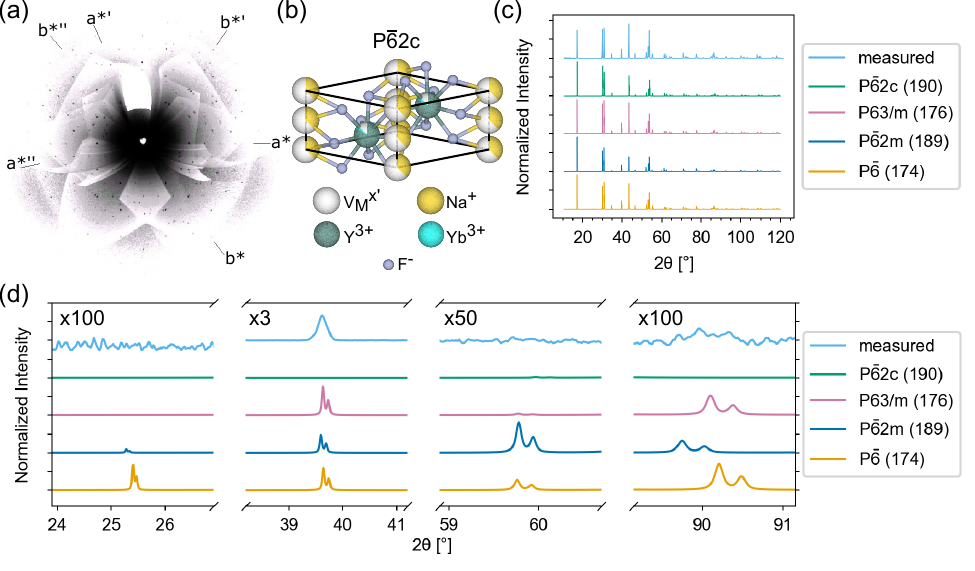}
    \centering
	\caption{(hk0) Burger projection calculated from the single crystal X-ray diffraction images of the measured \ch{Na_{1.00}MIDA} \ch{\textbeta-NaYF} drilling (a) and packing of the resulting refined crystal structure with P$\bar{6}$2c space group at \SI{100}{K} (b). Comparison of experimental and predicted powder X-ray diffraction patterns for published crystal structures in full view (c) and zoomed in for crucial diffractions (d).}
	\label{figure:XRD}
\end{figure}

\begin{figure}[t]
	\includegraphics[width=0.9\linewidth]{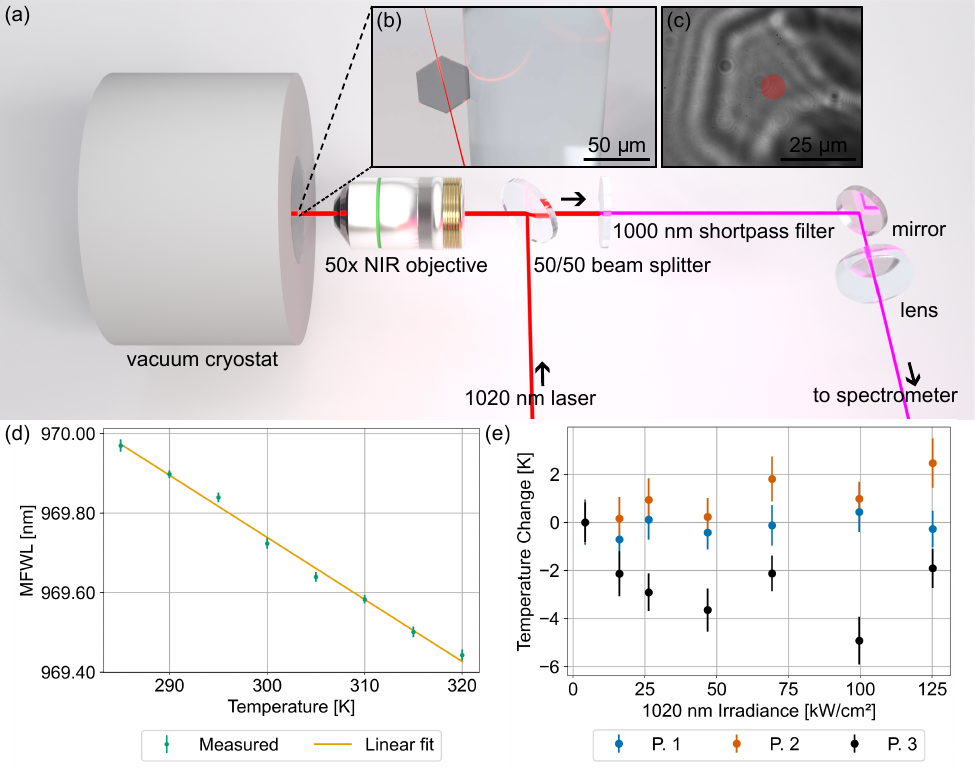}
    \centering
	\caption{Schematic of the experimental setup used for the mean fluorescence wavelength temperature calibration and laser power dependent laser refrigeration trials (a). Illustration (b) and optical image (c) of a \ch{\textbeta-NaYF} disk mounted on a fiber core to minimize substrate interactions. The illumination spot, with a size of  \SI{5}{\um}, is virtually overlaid as a red circle. 
    Measured temperature calibration curve (d) and \SI{1020}{nm} power-dependent temperature change of three different particles (b).}
	\label{figure:Optics}
\end{figure}

The size of the \ch{\textbeta-NaYF} hexagonal disks synthesized here is large enough to enable single-crystal X-ray diffraction measurements. The particle investigated was observed to be a triplet with rotational twinning around the z-axis of \SI{114.4}{\degree} and \SI{171.3}{\degree}. 
The measured diffraction pattern best matched the P$\bar{6}$2c (\#190) space group after refinement (see \textbf{Figure \ref{figure:XRD}} (a) and (b)). The sodium ions appear to be disordered on the c-axis with two locations at [000] and [000.5]. To the best of our knowledge, this structure has not been reported for \ch{\textbeta-NaYF} and is of higher symmetry than the commonly discussed P$\bar{6}$ (\#174), P$\bar{6}$2m (\#189) and P$6_3$m (\#176) space groups \cite{Wang2006, Fedorov2008, Liu2014a, Bessoi2016, Kraemer2004, Gai2012}.
Following the notation from Krämer et al. (2004)\cite{Kraemer2004} we found a composition of \ch{Na_{0.75}M_{1.75}F_6} (\ch{Na_{3x}M_{2-x}F_6} with $\textnormal{x}=0.25$) and M constituted of \SI{94.3}{\percent} \ch{Y^{3+}} and \SI{5.7} {\percent} \ch{Yb^{3+}}.
Complementary X-ray fluorescence (XRF) measurements showed a doping concentration of \SI{8.5(4)}{\percent} and detected no other lanthanides in the sample. A representative XRF spectrum is presented in \textbf{Figure S5}.
Predicted powder XRD patterns and experimentally obtained data are presented in panels (c) and (d) of Figure \ref{figure:XRD}. It becomes apparent that the observed diffraction peak around \SI{40}{\degree} 2\textTheta~is not predicted for the P$\bar{6}$2c space group, this is due to the fact that the corresponding (111) diffraction has a structure factor of zero caused by symmetry. The sodium ion disorder is only taken into account by average lattice occupancies and could potentially be the reason for a non-zero (111) diffraction intensity. The (111) diffraction is captured by the other published space groups (P$\bar{6}$, P$\bar{6}$2m and P$6_3$m), but unlike the proposed P$\bar{6}$2c space group, these also predict diffraction peaks that are not experimentally observed as shown in Figure \ref{figure:XRD} (d). The measured powder XRD data clearly show texturing and preferred orientation effects that distort relative intensities and, unlike the single crystal data, represent an ensemble average containing phase impurities, which could explain the observed diffraction around \SI{40}{\degree} 2\textTheta. These phase impurities could be caused by the non-stoichiometry and local disturbances of the \ch{\textbeta-NaYF} crystal structure. Because of the large surface area of the crystals, surface effects, like reconstructions, could alter the symmetry as well. Hypothetically \ch{\textbeta-NaYF} could exist in different crystal structures depending on the non-stoichiometry index $x$, which might be altered during growth, for example due to depletion of the precursor solution. Different synthetic conditions could further enhance the effects and different methods (hydrothermal, thermal coprecipitation, or thermal decomposition) even more so. This could be one of the reasons why several different crystal structures have been reported for \ch{\textbeta-NaYF}.

Beyond controlling the size and morphology, the synthesis was designed to maintain the laser refrigeration capabilities of \ch{\textbeta-NaYF} through substitutional doping with \ch{Yb^{3+}}. In order to assess the temperature change of the \ch{Na_{1.00}MIDA} \ch{\textbeta-NaYF} disks, power dependent anti-Stokes photoluminescence spectroscopy was performed on six different single crystals after calcination in air. A typical spectral response and the integration range for the mean fluorescence wavelength (MFWL) analysis are shown in \textbf{Figure S6}. Temperature calibration of the MFWL was achieved by measuring a particle at the lowest irradiance at different temperatures in a cryostat, the linear fit coefficients are \SI{974.4(2)}{nm} and \SI{-15.6(5)}{pm.K^{-1}} (see \textbf{Figure \ref{figure:Optics}}).
It should be noted that due to experimental limitations (detection filter cutoff at \SI{1000}{nm} the MFWL is underestimated. Nonetheless, the measured value accurately tracks the temperature.
Representative results of the power-dependent measurements are shown in Figure \ref{figure:Optics} (e). While two particles showed slight heating under increasing \SI{1020}{nm} irradiance (\SI{1.5(8)}{K} and \SI{2.5(10)}{K}), three particles showed no significant temperature change (\SI{-0.1(9)}{K}, \SI{0.2(7)}{K} and \SI{-0.3(8)}{K}) and one particle cooled by \SI{-4.9(10)}{K}. 
These results were obtained without any optimization of the dopant concentration or postsynthesis treatments like shelling, which have been shown to be able to greatly enhance quantum yields in upconversion nanoparticles and cooling efficiencies of \ch{\textbeta-NaYF} nanoparticles \cite{Yi2007, Haase2011, Fischer2016, Wurth2018, Laplane2024}.
The low cooling efficiency and percentage of refrigerating particles observed here are also seen for \ch{\textbeta-NaYF} hydrothermally synthesized using \ch{EDTA} and can potentially be attributed to the incorporation of \ch{OH-} ions into the host lattice. Thermal decomposition and thermal coprecipitation, the other common synthesis methods, lead to better laser refrigeration and optical properties \cite{Laplane2024}, but cannot readily produce microparticles with sizes and shapes designed for specific applications.
The particle-to-particle differences observed during the laser refrigeration trials are confirmed by the inhomogeneous broadening observed at \SI{4}{K} for an ensemble of particles presented in \textbf{Figure S7}. Further, the presence of at least five peaks, with different mono- or biexponential PL decay rates (\textbf{Figure S8}), indicates different local environments of the \ch{Yb^{3+}} dopant. This could be caused by different non-stoichiometries, crystal structures or Yb-ion interactions with adjacent dopants or surface defects. The measured integrated intensity drop from \SI{4}{K} to \SI{298}{K} indicates the presence of increasing nonradiative decay rates and underline the problems of optical quality for the hydrothermal synthesis presented here. 
However, some individual \ch{\textbeta-NaYF} particles show sufficient quality to achieve laser refrigeration. 
Further studies investigating the defect and crystal structure as well as non-stoichiometry might allow for an improvement in the laser refrigeration efficiency of hydrothermally grown \ch{\textbeta-NaYF} microparticles.

\section{Conclusion}

The substitution of ethylenediaminetetraacetic acid (\ch{EDTA}) with methyliminodiactetic acid (\ch{MIDA}) in hydrothermal synthesis allows for the growth of 
hexagonal \textbeta-NaYF disks with higher aspect ratios (\num{44}) and larger sizes 
(\SI{44.0}{\micro m} $\times$ \SI{1.0}{\micro m}) than those previously reported (\SI{5}{\micro m} $\times$ $\sim$\SI{500}{nm}) 
resulting in an aspect ratio of \num{10.5} \cite{Felsted2022}. While different protonation states of \ch{MIDA} allow for the synthesis of rods and novel semicircular disks, the adjustment ranges are small and byproducts like \ch{\textalpha-NaYF} and \ch{YF3} are present. The surface roughness of the disks obtained with \ch{MIDA} is comparable to those synthesized with \ch{EDTA}, but the \ch{MIDA} microparticles have flat basal planes, unlike the lensing observed with \ch{EDTA}. Single crystal XRD results indicate a P$\bar{6}$2c (\#190) space group of the \SI{44.0}{\micro m} \ch{\textbeta-NaYF} particles, which has not been reported in the literature. Deviations of experimentally obtained powder diffraction data are attributed to non-stoichiometry and local distortions. The crystal structure and non-stoichiometry of \ch{\textbeta-NaYF} is hypothesized to depend on specific synthesis conditions and processing parameters.
Laser refrigeration was observed with temperature decreases of \SI{-4.9(10)}{K} based on the anti-Stokes photoluminescence of \ch{Yb^{3+}}-dopants in one of six particles. The large aspect ratio of these microcrystals makes them promising candidates for future applications in optically levitated gravitational wave sensing and microoptics.

%%% Experimental section
\section{Experimental Section}
\textbf{Synthesis of \textbeta-\ch{NaYF}}\quad
\ch{YCl3 * 6 H2O} (\SI{99.9}{\percent}), \ch{YbCl3 * 6 H2O} 
(\SI{99.998}{\percent}), methyliminodiactetic acid (MIDA, \SI{99}{\percent}) and ethylenediaminetetraacetic acid (EDTA, \SI{99.995}{\percent}) were purchased from Sigma-Aldrich. \ch{NaOH} (\SI{99.5}{\percent}) was 
purchased from Fisher Scientific and NaF (\SI{99.5}{\percent}) was obtained from EMD Chemicals. 
Ultrapure water was used for all experiments (Barnstead GenPure). 
In a typical synthesis \SI{4}{\mL} of a \SI{0.25}{\Molar} aqueous rare-earth solution (\SI{90}{\percent} \ch{Y} and \SI{10}{\percent} \ch{Yb}) were chelated 
with \SI{5}{\mL} MIDA solution in an autoclave liner for \SI{10}{\min} while stirring. The protonation state of MIDA was controlled by dissolving \SI{2}{\mmol} in a total volume of \SI{5}{\mL} of a mixture of \ch{H2O} and \SI{5}{\Molar} \ch{NaOH}. 
\SI{5}{\mL} of \SI{0.8}{\Molar} \ch{NaF} solution was added, and the solution left to nucleate for \SI{30}{\min} under stirring. After the stir bar was removed, the liner was placed in an autoclave (Parr Instruments 4747) and a preheated oven at \SI{220}{\degreeCelsius} for \SI{72}{\hour}. After the autoclave naturally cooled to room temperature, the precipitate was collected and washed three times with \SI{30}{\mL} of \ch{H2O} and \ch{EtOH} respectively. The resulting product was dried at roughly \SI{80}{\degreeCelsius} overnight. 
Synthesis using \ch{EDTA} followed the same procedure, \SI{2}{\mmol} \ch{MIDA} were replaced by 
\SI{1}{\mmol} \ch{EDTA}.
If a mixture of \ch{\textalpha-NaYF} and 
\ch{\textbeta-NaYF} was present, the \ch{\textalpha-NaYF} particles were removed by dispersion in 
\ch{EtOH} followed by supernatant removal for three times.  Finally, the powder was calcined at 
\SI{300}{\degreeCelsius} for \SI{4}{\hour} in air to remove surface ligands and adsorbents. 

\textbf{Host Characterization}\quad
Scanning electron microscope (SEM) images were obtained on a Thermo Fisher Scientific Apreo 2 S using an accelerating voltage of \SI{2}{kV} and current of \SI{13}{pA}. 
Powder X-ray diffraction (XRD) data were obtained on a Bruker D8 Discover instrument. 
X-ray fluorescence was measured on a Bruker M4 Tornado equipped with a rhodium source operated at \SI{50}{keV} and \SI{600}{\micro A}. Quantitative data analysis of lanthanide ion concentrations was obtained with the corresponding Bruker software.
Atomic force microscopy (AFM) images were collected on a Dimension Icon AFM (Bruker) in tapping mode using Multi75Al probes (Budget Sensors). All images were recorded in air at room temperature. Offline data processing and analysis were done using Gwyddion software.
Single-crystal XRD measurements were conducted on a Bruker APEX II diffractometer using molybdenum (\ch{Mo}) radiation and an optical Miracol X-ray collimator. 
% A colorless prism, measuring $0.040 \times 0.035 \times 0.002$~\si{mm^3} 
The sample was mounted on a loop with oil.  Data were collected at \SI{-173}{\degreeCelsius}. The crystal-to-detector distance was 40 mm and exposure time was 60 seconds per frame for all sets using a scan width of \SI{1}{\degree}. More information on data processing and analysis can be found in the supporting information.

\textbf{Optical Characterization}\quad
The \SI{1020}{nm} laser power-dependent temperature change of single particles was evaluated using mean fluorescence wavelength thermometry \cite{LuntzMartin2021}, based on temperature-dependent spectral changes very similar to two-band differential luminescence thermometry \cite{Patterson2010}. The samples were loaded onto an optical fiber core, mounted in a liquid nitrogen cooled vacuum cryostat and irradiated through a \SI{50}{x} 
NIR-corrected microscope objective (50X Mitotuyo Plan Apo NIR Infinity Corrected) with a 
\SI{1020}{\nm} laser diode (QPhotonics, QDFBLD-1020-400). The optical response was measured 
by an EM-CCD (BLAZE 100-HR, Princeton Instruments) in an epifluorescence setup after passing a \SI{1000}{\nm} short-pass filter (see Figure \ref{figure:Optics}).

Temperature-dependent photoluminescence measurements were conducted on powder samples of the material prepared as a mull suspension by mixing dry powder with polydimethylsiloxane then sandwiched between two sapphire disks. The sample was loaded into a closed-cycle helium cryostat and measured under high vacuum (\SI{1e-6}{Pa}). Sample emission was focused into a monochromator with a spectral bandwidth of \SI{0.1}{nm} for spectral data and \SI{1}{nm} for time-resolved measurements. Spectra were collected using a liquid nitrogen-cooled silicon CCD camera and were corrected for instrument response. Time-resolved data was collected using a liquid nitrogen-cooled Hamamatsu InGaAs/InP NIR photomultiplier tube. Photoexcitation was performed using a tunable Ti:Sapphire laser at \SI{942}{nm}, \SI{150}{mW.cm^{-2}}, continuous-wave. For time-resolved experiments, the laser was at \SI{942}{nm}, \SI{500}{nJ.cm^{-2}}, \SI{2}{kHz}, \SI{1}{\us} pulse duration.

$\,$

$\,$

\printbibliography

%%%%%%%%%% Merge with supplemental materials %%%%%%%%%%
\pagebreak
% \widetext
\begin{center}
\textbf{\large Supplemental Materials: Hydrothermal Synthesis of Ultra-high Aspect Ratio \textbeta-NaYF Disks via Methyliminodiacetic Acid (MIDA)}
\end{center}
%%%%%%%%%% Merge with supplemental materials %%%%%%%%%%
%%%%%%%%%% Prefix a "S" to all equations, figures, tables and reset the counter %%%%%%%%%%
\setcounter{equation}{0}
\setcounter{figure}{0}
\setcounter{table}{0}
\setcounter{page}{1}
\makeatletter
\renewcommand{\theequation}{S\arabic{equation}}
\renewcommand{\thefigure}{S\arabic{figure}}
\renewcommand{\thetable}{S\arabic{table}}
\newrefsection
%%%%%%%%%% Prefix a "S" to all equations, figures, tables and reset the counter %%%%%%%%%%

\begin{figure}[htb!]
	\includegraphics[width=0.9\linewidth]{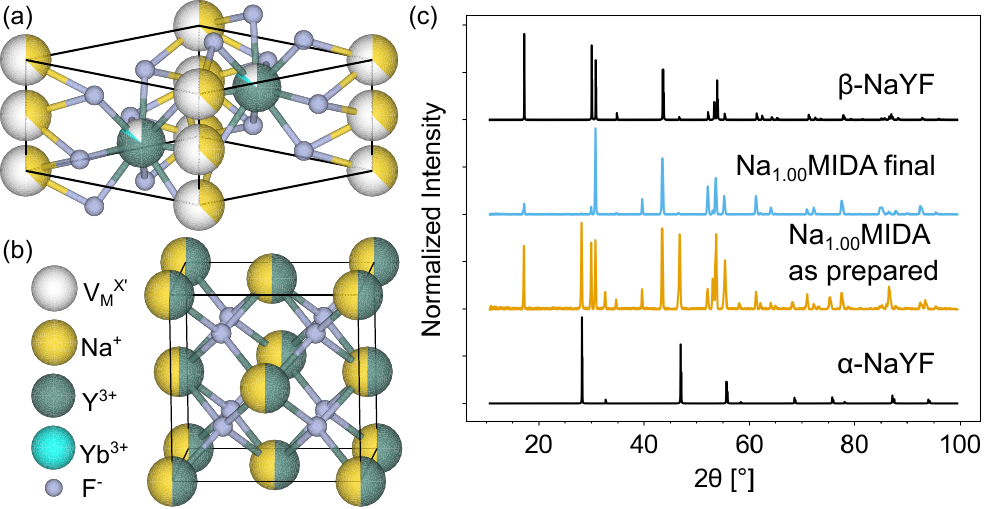}
    \centering
	\caption{Crystal structure of \textbeta- (a) and \textalpha-NaYF (b) \cite{Wang2006}. XRD patterns (c) of 
		\ch{Na_{1.00}MIDA}  synthesis before and after \textalpha-NaYF removal and calcination.}
	\label{figure:XRD}
\end{figure}

\begin{figure}[htb!]
	\includegraphics[width=0.9\linewidth]{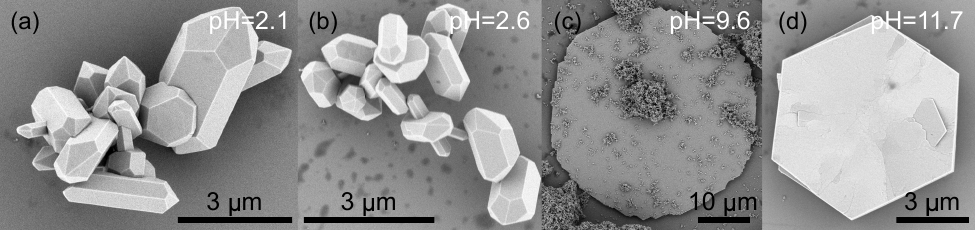}
	\centering
    \caption{Hydrothermal synthesis results using MIDA in different protonation states by the addition 
		of \SI{0.00}{\eq} (a), \SI{0.50}{\eq} (b), \SI{1.25}{\eq} (c) and \SI{2.00}{\eq} (d).}
\end{figure}

\begin{figure}[htb!]
	\includegraphics[width=0.9\linewidth]{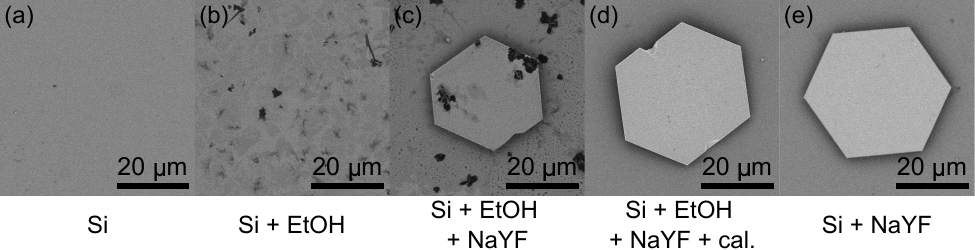}
	\centering
    \caption{SEM images of cleaned Si substrate (a), Si substrate after drop casting ethanol (b), \ch{MIDA} \ch{\textbeta-NaYF} drop cast from ethanol (c), drop cast from ethanol and calcined (d) and dry attached to the Si substrate (e).}
\end{figure}

\begin{figure}[htb!]
	\includegraphics[width=0.9\linewidth]{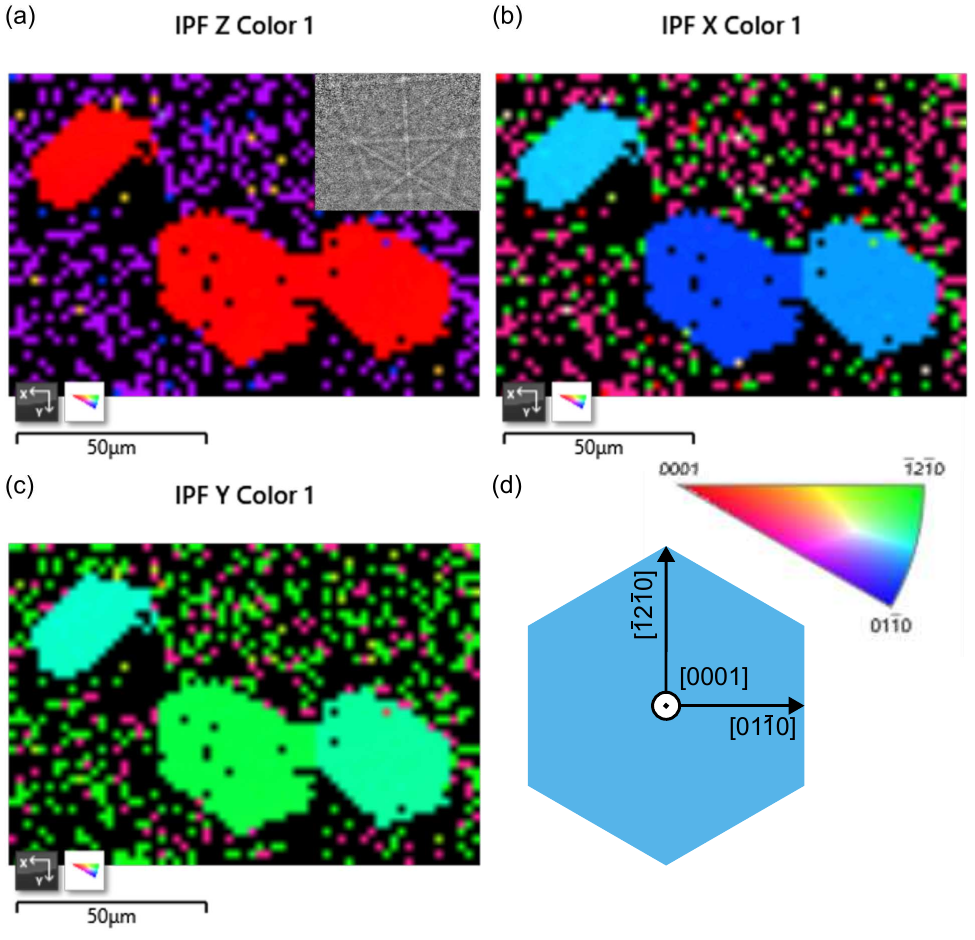}
	\centering
    \caption{Polar plots of \ch{Na_{1.00}MIDA} \ch{\textbeta-NaYF} obtained with electron backsactter diffraction (EBSD) (a)-(c) and crystallographic orientation of the hexagonal particles (d). The inset in (a) shows a representative single point Kikuchi pattern.}
\end{figure}

\clearpage

\textbf{Single crystal XRD data processing and analysis}\quad
Data collection was \SI{96.1}{\percent} complete to \SI{25}{\degree} in \texttheta.  A total of  250 merged reflections were collected covering the indices, $-7<=h<=7$, $-8<=k<=9$, $-5<=l<=5$. 137 reflections were symmetry independent and the $R_{int} = 0.0166$ indicated that the data was of excellent quality (average \num{0.07}). Indexing and unit cell refinement indicated a primitive hexagonal lattice.  The space group was found to be P$\bar{6}$2c (\#190). 

The data appeared to be of a drilling (CELL\_NOW \cite{Sheldrick2005}) and after multi-domain integration (SAINT, SADABS within the APEX2 software package by Bruker \cite{Bruker2007}) the data was merged utilizing TWINABS \cite{Sheldrick2007_TWINABS}.

Solutions by direct methods (SHELXT \cite{Sheldrick2007_SHELX, Sheldrick2015} or SIR97 \cite{Altomare1993, Altomare1999}) produced a complete heavy atom phasing model consistent with the proposed structure. The structure was completed by difference Fourier synthesis with SHELXL \cite{Sheldrick1997, Sheldrick2015a}. Scattering factors are from Waasmair and Kirfel \cite{Waasmaier1995}. All atoms were refined anisotropically by full-matrix least-squares.

The sodium appears disordered on the c-axis with two locations at [0 0 0] and [0 0 0.5] (symmetry related to the actual coordinates in the refinement). The sodium site apperently did not contain any yttrium and vice versa. The occupancies were fitted from satisfying equation 6 = occup.(Na) + 3*occup.(Y) to ensure the sum of the cations chareges adds up to +6. The space group is non-centrosymmetric but the sample appears to be a racemate with Flack parameter of 0.5(7). No higher symmetry was found.

Table \ref{table:SXRD} summarizes the data collection details.
\bigskip

\begin{table}[tb]
	\caption{Crystallographic data for the \ch{Na_{1.00}MIDA} \ch{\textbeta-NaYF} structure provided.}
    \label{table:SXRD}
    \centering
	\begin{tabular}{ll}
		%\hline
		Empirical formula & \ch{F6Na_{0.75}Y_{1.65}Yb_{0.10}} \\
        Formula weight & 295.25 g mol\textsuperscript{-1}  \\
        Temperature & 100(2) K  \\
        Wavelength & \SI{0.71073}{\angstrom}  \\
        Crystal system & Hexagonal \\
        Space group & P -6 2 c \\
        Unit cell dimensions & a = 5.9451(5) \unit{\angstrom} \qquad \textalpha\ = \SI{90}{\degree}\\
         & b = 5.9451(5) \unit{\angstrom} \qquad \textbeta\  = \SI{90}{\degree}\\
         & c = 3.5054(4) \unit{\angstrom} \qquad \textbeta\  = \SI{120}{\degree}\\
         Volume & 107.30(2) \unit{\angstrom^3}\\ 
         Z & 1\\
         Density (calculated) & 4.569 Mg m\textsuperscript{-3}\\
         Absorption coefficient & 24.512 mm\textsuperscript{-1}\\
         F(000) & 134\\
         Crystal size & 0.040 x 0.035 x 0.002 mm\textsuperscript{3}\\
         Theta range for data collection & 3.958 to \SI{32.776}{\degree}\\ 
         Index ranges & -7<=h<=7, -8<=k<=9, -5<=l<=5\\ 
         Reflections collected & 250\\ 
         Independent reflections & 137 [R(int) = 0.0166]\\ 
         Completness to thetha = \SI{25.242}{\degree} & \SI{96.1}{\percent}\\ 
         Refinement method & Full-matrix least-squares on F\textsuperscript{2}\\ 
         Data / restraints / parameters & 137 / 1 / 15\\ 
         Goodness-of-fit on F\textsuperscript{2} & 1.396\\ 
         Final R indices [I>2sigma(I)] & R1 = 0.0318, wR2 = 0.0863\\
         R indices (all data) & R1 = 0.0319, wR2 = 0.0863\\
         Absolute structure parameter & 0.51(7)\\ 
         Largest diff. peak and hole & 1.489 and -1.177 e.\unit{\angstrom^3}\\
		%\hline
	\end{tabular}
\end{table}

\clearpage

\begin{figure}[htb!]
	\includegraphics[width=0.9\linewidth]{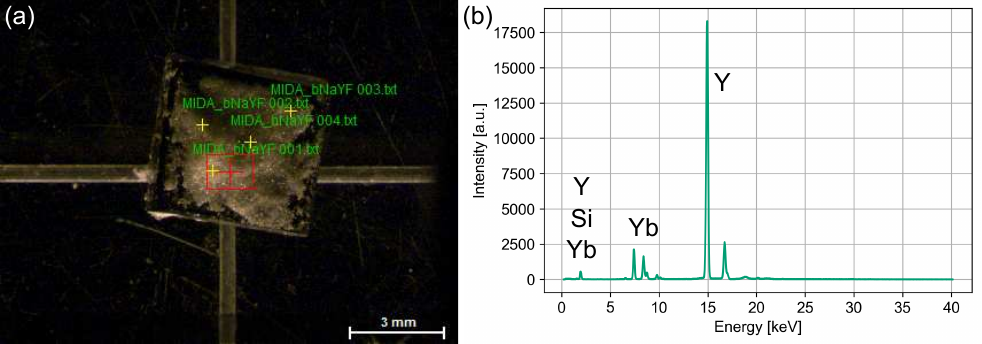}
	\centering
    \caption{Microscope image of \ch{Na_{1.00}MIDA} \ch{\textbeta-NaYF} drop cast from EtOH on SI (a) and representative X-ray fluorescence spectrum of the sample (b). The average \ch{Yb^{3+}} doping concentration from four point measurements was \SI{8.5(4)}{\percent}.}
\end{figure}

% \begin{table}[tb]
% 	\centering
%     \caption{Relative yttrium and ytterbium concentrations measured for \ch{Na_{1.00}MIDA} \ch{\textbeta-NaYF} sample using X-ray fluorescence (XRF). No other lanthanide ions were detected.}
% 	\label{table:NaOHOptimizationResults}
% 	\begin{tabular}{lcc}
% 		\hline
% 		Spot & Y [at\%] & Yb [at\%] \\
% 		\hline
% 		1 & \num{91.8} & \num{8.2} \\
%         2 & \num{90.1} & \num{9.9} \\
%         3 & \num{93.0} & \num{7.0} \\
%         4 & \num{91.5} & \num{8.5} \\
% 		\hline
% 	\end{tabular}
% \end{table}

\begin{figure}[htb!]
	\includegraphics[width=0.9\linewidth]{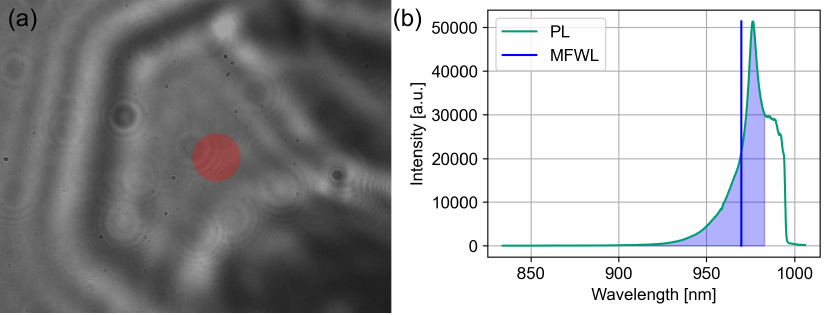}
    \centering
	\caption{Image of a \ch{Na_{1.00}MIDA} \ch{\textbeta-NaYF} particle on a fiber core with the \SI{1020}{nm} focussed laser spot indicated in red (a). Spectral anti-Stokes response of the \ch{Yb^{3+}} dopant and mean fluorescence wavelength (MFWL) integration range (b).}
\end{figure}

\begin{figure}[htb!]
	\includegraphics[width=0.9\linewidth]{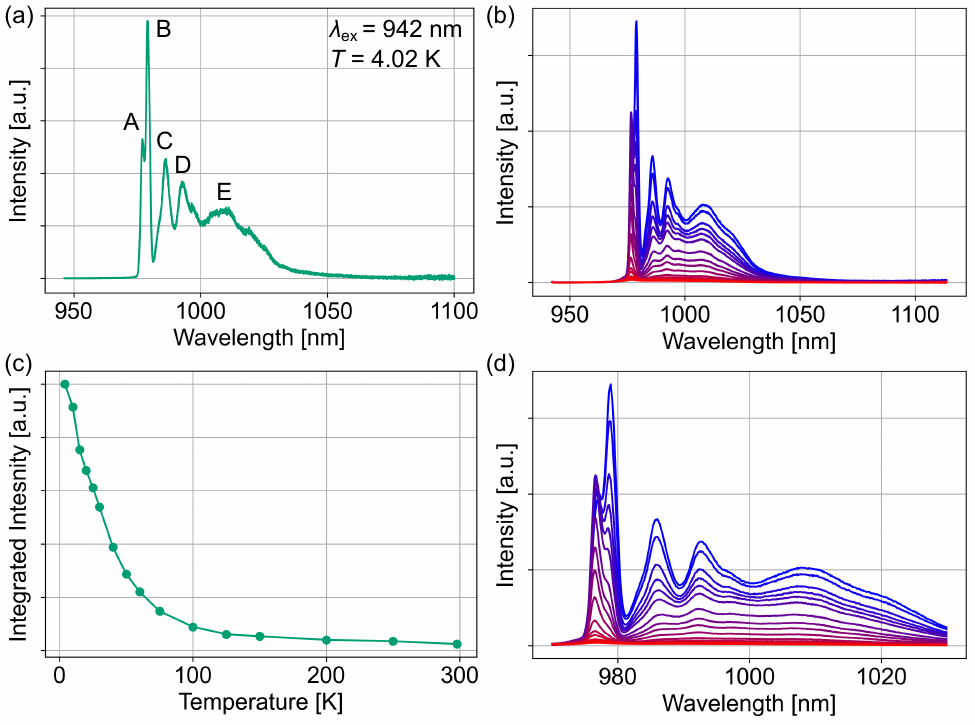}
    \centering
	\caption{Low temperature PL spectrum of an ensemble of \ch{Na_{1.00}MIDA} \ch{\textbeta-NaYF} particles under \SI{942}{nm} continuous wave excitation (a). Variable temperature PL from \SI{4}{K} to \SI{298}{K} (blue to red) (b). Temperature-dependent integrated PL intensity over the entire spectral region (c). Zoomed in version of variable temperature PL (d).}
\end{figure}

\begin{figure}
\begin{floatrow}
\ffigbox{%
  \centering
  \includegraphics[width=0.45\textwidth]{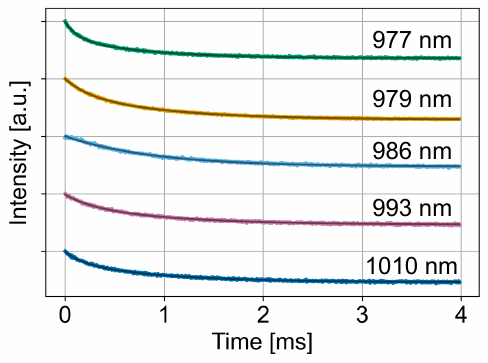}%
}{%
  \caption{PL decay plots of the A, B, C, D and E peaks labeled in Figure S7, excited using a \SI{942}{nm} source with \SI{2000}{Hz} frequency and \SI{1}{\micro s} pulse. Black curves are fit using a mono- (C) or biexponential decay functions (A, B, D, E).}%
}
\ttabbox{%
  \centering
  \begin{tabular}{ccc}
		\hline
		\textlambda\ [\si{nm}] & $\tau_{1}$ [\si{ms}] & $\tau_{2}$ [\si{ms}] \\
		\hline
		\num{977} & \num{0.765} (\SI{72}{\percent}) & \num{0.128} (\SI{28}{\percent}) \\
        \num{979} & \num{0.940} (\SI{82}{\percent}) & \num{0.229} (\SI{18}{\percent}) \\
        \num{986} & \num{0.895} (\SI{100}{\percent}) & N/A \\
        \num{993} & \num{0.991} (\SI{83}{\percent}) & \num{0.263} (\SI{17}{\percent})  \\
        \num{1010} & \num{0.948} (\SI{82}{\percent}) & \num{0.197} (\SI{18}{\percent})  \\
		\hline
	\end{tabular}
}{%
  \caption{Fitting results using mono- or biexponetial decays for the time-resolved photoluminescence of peaks A-E.
  Relative integrated intensities are given in brackets.}%
}
\end{floatrow}
\end{figure}

%%% SI References
\medskip
\clearpage

\printbibliography

@article{Aebischer2006,
  title = {Structural and {{Spectroscopic Characterization}} of {{Active Sites}} in a {{Family}} of {{Light}}-{{Emitting Sodium Lanthanide Tetrafluorides}}},
  author = {Aebischer, Annina and Hostettler, Marc and Hauser, J{\"u}rg and Kr{\"a}mer, Karl and Weber, Thomas and G{\"u}del, Hans Ulrich and B{\"u}rgi, Hans-Beat},
  year = 2006,
  month = apr,
  journal = {Angewandte Chemie International Edition},
  volume = {45},
  number = {17},
  pages = {2802--2806},
  publisher = {Wiley},
  issn = {1521-3773},
  doi = {10.1002/anie.200503966}
}

@article{Aggarwal2022,
  title = {Searching for {{New Physics}} with a {{Levitated-Sensor-Based Gravitational-Wave Detector}}},
  author = {Aggarwal, Nancy and Winstone, George P. and Teo, Mae and Baryakhtar, Masha and Larson, Shane L. and Kalogera, Vicky and Geraci, Andrew A.},
  year = 2022,
  month = mar,
  journal = {Physical Review Letters},
  volume = {128},
  number = {11},
  pages = {111101},
  publisher = {American Physical Society (APS)},
  issn = {1079-7114},
  doi = {10.1103/physrevlett.128.111101}
}

@article{Altomare1993,
  title = {Completion and Refinement of Crystal Structures {{withSIR92}}},
  author = {Altomare, A. and Cascarano, G. and Giacovazzo, C. and Guagliardi, A.},
  year = 1993,
  month = jun,
  journal = {Journal of Applied Crystallography},
  volume = {26},
  number = {3},
  pages = {343--350},
  publisher = {International Union of Crystallography (IUCr)},
  issn = {0021-8898},
  doi = {10.1107/s0021889892010331}
}

@article{Altomare1999,
  title = {{{SIR97}}: A New Tool for Crystal Structure Determination and Refinement},
  author = {Altomare, Angela and Burla, Maria Cristina and Camalli, Mercedes and Cascarano, Giovanni Luca and Giacovazzo, Carmelo and Guagliardi, Antonietta and Moliterni, Anna Grazia Giuseppina and Polidori, Giampiero and Spagna, Riccardo},
  year = 1999,
  month = feb,
  journal = {Journal of Applied Crystallography},
  volume = {32},
  number = {1},
  pages = {115--119},
  publisher = {International Union of Crystallography (IUCr)},
  issn = {0021-8898},
  doi = {10.1107/s0021889898007717}
}

@article{Bessoi2016,
  title = {Rapid Microwave Mediated Hydrothermal Synthesis of Complex Ternary Fluorides},
  author = {Bessoi, Monalisa and Soren, Siba and Parhi, Purnendu},
  year = 2016,
  month = feb,
  journal = {Ceramics International},
  volume = {42},
  number = {2},
  pages = {3697--3700},
  publisher = {Elsevier BV},
  issn = {0272-8842},
  doi = {10.1016/j.ceramint.2015.10.149}
}

@article{Bogachev2022,
  title = {Lanthanide-{{Ion-Doping Effect}} on the {{Morphology}} and the {{Structure}} of {{NaYF4}}:{{Ln3}}+ {{Nanoparticles}}},
  shorttitle = {Lanthanide-{{Ion-Doping Effect}} on the {{Morphology}} and the {{Structure}} of {{NaYF4}}},
  author = {Bogachev, Nikita A. and Betina, Anna A. and Bulatova, Tatyana S. and Nosov, Viktor G. and Kolesnik, Stefaniia S. and Tumkin, Ilya I. and Ryazantsev, Mikhail N. and Skripkin, Mikhail Yu. and Mereshchenko, Andrey S.},
  year = 2022,
  month = aug,
  journal = {Nanomaterials},
  volume = {12},
  number = {17},
  pages = {2972},
  issn = {2079-4991},
  doi = {10.3390/nano12172972},
  urldate = {2025-05-19},
  copyright = {https://creativecommons.org/licenses/by/4.0/},
  langid = {english}
}

@misc{Bruker2007,
  title = {{{APEX2}} ({{Version}} 2.1-4), {{SAINT}} (Version 7.{{34A}}), {{SADABS}} (Version 2007/4)},
  author = {{Bruker}},
  year = 2007
}

@article{Chen2012,
  title = {{$\beta$}-{{NaYF4}}:{{Er3}}+(10\%) Microprisms for the Enhancement of a-{{Si}}:{{H}} Solar Cell near-Infrared Responses},
  shorttitle = {{$\beta$}-{{NaYF4}}},
  author = {Chen, Yongsheng and He, Wei and Jiao, Yuechao and Wang, Honghong and Hao, Xiuli and Lu, Jingxiao and Yang, Shi'e Er},
  year = 2012,
  month = sep,
  journal = {Journal of Luminescence},
  volume = {132},
  number = {9},
  pages = {2247--2250},
  issn = {00222313},
  doi = {10.1016/j.jlumin.2012.04.011},
  urldate = {2025-05-15},
  copyright = {https://www.elsevier.com/tdm/userlicense/1.0/},
  langid = {english}
}

@article{Chen2015a,
  title = {Enhanced High-Order Upconversion Luminescence of Hexagonal Phase {{NaYF}} 4 :{{Yb}} 3+ ,{{Tm}} 3+ Crystals Coated with Homogeneous Shell},
  shorttitle = {Enhanced High-Order Upconversion Luminescence of Hexagonal Phase {{NaYF}} 4},
  author = {Chen, Huan and Lang, Yanbo and Zhao, Dan and He, Chunfeng and Qin, Weiping},
  year = 2015,
  month = jun,
  journal = {Journal of Fluorine Chemistry},
  volume = {174},
  pages = {70--74},
  issn = {00221139},
  doi = {10.1016/j.jfluchem.2015.02.019},
  urldate = {2025-05-19},
  langid = {english}
}

@article{Chen2020,
  title = {Recent Advances in the Synthesis and Application of {{Yb-based}} Fluoride Upconversion Nanoparticles},
  author = {Chen, Bing and Wang, Feng},
  year = 2020,
  journal = {Inorganic Chemistry Frontiers},
  volume = {7},
  number = {5},
  pages = {1067--1081},
  publisher = {Royal Society of Chemistry (RSC)},
  doi = {10.1039/c9qi01358j}
}

@article{Fedorov2008,
  title = {Soft Chemical Synthesis of {{NaYF4}} Nanopowders},
  author = {Fedorov, P. P. and Kuznetsov, S. V. and Voronov, V. V. and Yarotskaya, I. V. and Arbenina, V. V.},
  year = 2008,
  month = nov,
  journal = {Russian Journal of Inorganic Chemistry},
  volume = {53},
  number = {11},
  pages = {1681--1685},
  publisher = {Pleiades Publishing Ltd},
  issn = {1531-8613},
  doi = {10.1134/s0036023608110028}
}

@article{Felsted2022,
  title = {Chemically {{Tunable Aspect Ratio Control}} and {{Laser Refrigeration}} of {{Hexagonal Sodium Yttrium Fluoride Upconverting Materials}}},
  author = {Felsted, R. Greg and Pant, Anupum and Bard, Alexander B. and Xia, Xiaojing and {Luntz-Martin}, Danika R. and Dadras, Siamak and Zhang, Shuai and Vamivakas, A. Nick and Pauzauskie, Peter J.},
  year = 2022,
  month = may,
  journal = {Crystal Growth \& Design},
  volume = {22},
  number = {6},
  pages = {3605--3612},
  publisher = {American Chemical Society (ACS)},
  doi = {10.1021/acs.cgd.1c01174}
}

@article{FernandezBravo2018,
  title = {Continuous-Wave Upconverting Nanoparticle Microlasers},
  author = {{Fernandez-Bravo}, Angel and Yao, Kaiyuan and Barnard, Edward S. and Borys, Nicholas J. and Levy, Elizabeth S. and Tian, Bining and Tajon, Cheryl A. and Moretti, Luca and Altoe, M. Virginia and Aloni, Shaul and Beketayev, Kenes and Scotognella, Francesco and Cohen, Bruce E. and Chan, Emory M. and Schuck, P. James},
  year = 2018,
  month = jun,
  journal = {Nature Nanotechnology},
  volume = {13},
  number = {7},
  pages = {572--577},
  publisher = {{Springer Science and Business Media LLC}},
  issn = {1748-3395},
  doi = {10.1038/s41565-018-0161-8},
  comment-lars = {- 5 um PS microlasers with UCNPs - predecessor to Liu2020}
}

@article{Fischer2016,
  title = {Precise {{Tuning}} of {{Surface Quenching}} for {{Luminescence Enhancement}} in {{Core}}--{{Shell Lanthanide-Doped Nanocrystals}}},
  author = {Fischer, Stefan and Bronstein, Noah D. and Swabeck, Joseph K. and Chan, Emory M. and Alivisatos, A. Paul},
  year = 2016,
  month = nov,
  journal = {Nano Letters},
  volume = {16},
  number = {11},
  pages = {7241--7247},
  issn = {1530-6984, 1530-6992},
  doi = {10.1021/acs.nanolett.6b03683},
  urldate = {2025-05-20},
  langid = {english}
}

@article{Fu2017,
  title = {Enhanced Upconversion Luminescence of {{NaYF}} 4 :{{Yb}}, {{Er}} Microprisms via {{La}} 3+ Doping},
  shorttitle = {Enhanced Upconversion Luminescence of {{NaYF}} 4},
  author = {Fu, Junxiang and Zhang, Xiaozeng and Chao, Zhicong and Li, Zibo and Liao, Jinsheng and Hou, Dejian and Wen, Herui and Lu, Xiaoneng and Xie, Xinrong},
  year = 2017,
  month = feb,
  journal = {Optics \& Laser Technology},
  volume = {88},
  pages = {280--286},
  issn = {00303992},
  doi = {10.1016/j.optlastec.2016.09.029},
  urldate = {2025-05-19},
  langid = {english}
}

@article{Gai2012,
  title = {Facile Synthesis and Up-Conversion Properties of Monodisperse Rare Earth Fluoride Nanocrystals},
  author = {Gai, Shili and Yang, Guixin and Li, Xingbo and Li, Chunxia and Dai, Yunlu and He, Fei and Yang, Piaoping},
  year = 2012,
  journal = {Dalton Transactions},
  volume = {41},
  number = {38},
  pages = {11716},
  publisher = {Royal Society of Chemistry (RSC)},
  issn = {1477-9234},
  doi = {10.1039/c2dt30954h}
}

@article{Haase2011,
  title = {Upconverting {{Nanoparticles}}},
  author = {Haase, Markus and Sch{\"a}fer, Helmut},
  year = 2011,
  month = jun,
  journal = {Angewandte Chemie International Edition},
  volume = {50},
  number = {26},
  pages = {5808--5829},
  issn = {1433-7851, 1521-3773},
  doi = {10.1002/anie.201005159},
  urldate = {2025-05-09},
  copyright = {http://onlinelibrary.wiley.com/termsAndConditions\#vor},
  langid = {english}
}

@article{Hao2012,
  title = {Controlled Growth along Circumferential Edge and Upconverting Luminescence of {$\beta$}-{{NaYF4}}: 20\%{{Yb3}}+, 1\%{{Er3}}+ Microcrystals},
  shorttitle = {Controlled Growth along Circumferential Edge and Upconverting Luminescence of {$\beta$}-{{NaYF4}}},
  author = {Hao, Shuwei and Chen, Guanying and Qiu, Hailong and Xu, Chao and Fan, Rongwei and Meng, Xiangbin and Yang, Chunhui},
  year = 2012,
  month = nov,
  journal = {Materials Chemistry and Physics},
  volume = {137},
  number = {1},
  pages = {97--102},
  issn = {02540584},
  doi = {10.1016/j.matchemphys.2012.08.045},
  urldate = {2025-05-15},
  copyright = {https://www.elsevier.com/tdm/userlicense/1.0/},
  langid = {english}
}

@article{Hehlen2018,
  title = {First Demonstration of an All-Solid-State Optical Cryocooler},
  author = {Hehlen, Markus P. and Meng, Junwei and Albrecht, Alexander R. and Lee, Eric R. and Gragossian, Aram and Love, Steven P. and Hamilton, Christopher E. and Epstein, Richard I. and {Sheik-Bahae}, Mansoor},
  year = 2018,
  month = jun,
  journal = {Light: Science \& Applications},
  volume = {7},
  number = {1},
  pages = {15},
  issn = {2047-7538},
  doi = {10.1038/s41377-018-0028-7},
  urldate = {2025-11-10},
  langid = {english}
}

@article{Khare2020,
  title = {A Critical Review on the Efficiency Improvement of Upconversion Assisted Solar Cells},
  author = {Khare, Ayush},
  year = 2020,
  month = apr,
  journal = {Journal of Alloys and Compounds},
  volume = {821},
  pages = {153214},
  publisher = {Elsevier BV},
  doi = {10.1016/j.jallcom.2019.153214}
}

@inproceedings{Knowles1989,
  title = {Optical {{Properties}} of {{Nd}}:{{NaYF4}} and {{Ho}}:{{NaYF4}}},
  booktitle = {Advanced {{Solid State Lasers}}},
  author = {Knowles, David and Cassanho, A. and Jenssen, H. P.},
  year = 1989,
  series = {{{ASSL}}},
  volume = {70},
  pages = {CC7},
  publisher = {OSA},
  doi = {10.1364/assl.1989.cc7}
}

@article{Kraemer2004,
  title = {Hexagonal {{Sodium Yttrium Fluoride Based Green}} and {{Blue Emitting Upconversion Phosphors}}},
  author = {Kr{\"a}mer, Karl W. and Biner, Daniel and Frei, Gabriela and G{\"u}del, Hans U. and Hehlen, Markus P. and L{\"u}thi, Stefan R.},
  year = 2004,
  month = mar,
  journal = {Chemistry of Materials},
  volume = {16},
  number = {7},
  pages = {1244--1251},
  publisher = {American Chemical Society (ACS)},
  issn = {1520-5002},
  doi = {10.1021/cm031124o}
}

@article{Laplane2024,
  title = {Inert Shell Coating for Enhanced Laser Refrigeration of Nanoparticles: {{Application}} in Levitated Optomechanics},
  author = {Laplane, Cyril and Ren, Peng and Roberts, Reece P. and Lu, Yiqing and Volz, Thomas},
  year = 2024,
  month = mar,
  journal = {ACS Photonics},
  volume = {11},
  number = {3},
  pages = {963--968},
  publisher = {American Chemical Society (ACS)},
  issn = {2330-4022},
  doi = {10.1021/acsphotonics.3c01065}
}

@article{Li2007,
  title = {Different {{Microstructures}} of {$\beta$}-{{NaYF}}{\textsubscript{4}} {{Fabricated}} by {{Hydrothermal Process}}: {{Effects}} of {{pH Values}} and {{Fluoride Sources}}},
  shorttitle = {Different {{Microstructures}} of {$\beta$}-{{NaYF}}{\textsubscript{4}} {{Fabricated}} by {{Hydrothermal Process}}},
  author = {Li, Chunxia and Yang, Jun and Quan, Zewei and Yang, Piaoping and Kong, Deyan and Lin, Jun},
  year = 2007,
  month = oct,
  journal = {Chemistry of Materials},
  volume = {19},
  number = {20},
  pages = {4933--4942},
  issn = {0897-4756, 1520-5002},
  doi = {10.1021/cm071668g},
  urldate = {2025-05-14},
  langid = {english}
}

@article{Liu2014a,
  title = {Remarkable Enhancement of the Near-Infrared Upconversion Emission in the {$\beta$}-{{NaYF4}}:{{Yb3}}+/{{Tm3}}+ System with Controllable Morphology},
  author = {Liu, Bi-Qiu and Wang, Jiao and Zhu, Lin-Lin and Zhang, Hui and Chen, Hao-Hong and Xu, Fang-Fang and Guo, Kai and Zhao, Jing-Tai},
  year = 2014,
  month = mar,
  journal = {Materials Research Bulletin},
  volume = {51},
  pages = {180--184},
  publisher = {Elsevier BV},
  issn = {0025-5408},
  doi = {10.1016/j.materresbull.2013.11.057}
}

@article{Liu2020,
  title = {Controlled Assembly of Upconverting Nanoparticles for Low-Threshold Microlasers and Their Imaging in Scattering Media},
  author = {Liu, Yawei and Teitelboim, Ayelet and {Fernandez-Bravo}, Angel and Yao, Kaiyuan and Altoe, M. Virginia P. and Aloni, Shaul and Zhang, Chunhua and Cohen, Bruce E. and Schuck, P. James and Chan, Emory M.},
  year = 2020,
  month = feb,
  journal = {ACS Nano},
  volume = {14},
  number = {2},
  pages = {1508--1519},
  publisher = {American Chemical Society (ACS)},
  issn = {1936-086X},
  doi = {10.1021/acsnano.9b06102},
  comment-lars = {- 5 um PS with bNaYF UCNPs for lasing - basis for XX particles}
}

@article{LuntzMartin2021,
  title = {Laser Refrigeration of Optically Levitated Sodium Yttrium Fluoride Nanocrystals},
  author = {{Luntz-Martin}, Danika R. and Felsted, R. Greg and Dadras, Siamak and Pauzauskie, Peter J. and Vamivakas, A. Nick},
  year = 2021,
  month = jul,
  journal = {Optics Letters},
  volume = {46},
  number = {15},
  pages = {3797},
  publisher = {The Optical Society},
  doi = {10.1364/ol.426334}
}

@article{Malhotra2023,
  title = {Lanthanide-{{Doped Upconversion Nanoparticles}}: {{Exploring A Treasure Trove}} of {{NIR-Mediated Emerging Applications}}},
  author = {Malhotra, Karan and Hrovat, David and Kumar, Balmiki and Qu, Grace and Houten, Justin Van and Ahmed, Reda and Piunno, Paul A. E. and Gunning, Patrick T. and Krull, Ulrich J.},
  year = 2023,
  month = jan,
  journal = {ACS Applied Materials \& Interfaces},
  publisher = {American Chemical Society (ACS)},
  doi = {10.1021/acsami.2c12370}
}

@inproceedings{Meng2018,
  title = {Realization of an All-Solid-State Cryocooler Using Optical Refrigeration},
  booktitle = {Tri-{{Technology Device Refrigeration}} ({{TTDR}}) {{III}}},
  author = {Meng, Junwei and Gragossian, Aram and Lee, Eric and Ghasemkhani, Mohammadreza and Albrecht, Alexander R. and Volpi, Azzurra and Hehlen, Markus P. and Epstein, Richard I. and {Sheik-Bahae}, Mansoor},
  editor = {Epstein, Richard I. and Andresen, Bj{\o}rn F. and Benschop, Tonny and Heremans, Joseph P. and {Sheik-Bahae}, Mansoor and Riabzev, Sergey V.},
  year = 2018,
  month = may,
  pages = {10},
  publisher = {SPIE},
  address = {Orlando, United States},
  doi = {10.1117/12.2305195},
  urldate = {2025-11-10}
}

@article{Menyuk1972,
  title = {{{NaYF4}} : {{Yb}},{{Er}}---an Efficient Upconversion Phosphor},
  shorttitle = {{{NaYF4}}},
  author = {Menyuk, N. and Dwight, K. and Pierce, J.W.},
  year = 1972,
  month = aug,
  journal = {Applied Physics Letters},
  volume = {21},
  number = {4},
  pages = {159--161},
  issn = {0003-6951, 1077-3118},
  doi = {10.1063/1.1654325},
  urldate = {2025-05-09},
  langid = {english}
}

@article{Ockerbloom1956,
  title = {Thermodynamic {{Dissociation Constants}} of {{Methylaminediacetic Acid}} and {{Dimethylethylenediaminediacetic Acid}}},
  author = {Ockerbloom, Nelson E. and Martell, Arthur E.},
  year = 1956,
  month = jan,
  journal = {Journal of the American Chemical Society},
  volume = {78},
  number = {2},
  pages = {267--270},
  issn = {0002-7863, 1520-5126},
  doi = {10.1021/ja01583a004},
  urldate = {2025-05-28},
  langid = {english}
}

@article{OrtizRivero2021,
  title = {Laser Refrigeration by an Ytterbium-Doped {{NaYF}} {$_4$} Microspinner},
  author = {{Ortiz-Rivero}, Elisa and Prorok, Katarzyna and Mart{\'\i}n, Inocencio Rafael and Lisiecki, Rados{\l}aw and {Haro-Gonz{\'a}lez}, Patricia and Bednarkiewicz, Artur and Jaque, Daniel},
  year = 2021,
  month = sep,
  journal = {Small},
  volume = {17},
  number = {46},
  pages = {2103122},
  publisher = {Wiley},
  doi = {10.1002/smll.202103122}
}

@article{Patterson2010,
  title = {Measurement of Solid-State Optical Refrigeration by Two-Band Differential Luminescence Thermometry},
  author = {Patterson, W. M. and Seletskiy, D. V. and {Sheik-Bahae}, M. and Epstein, R. I. and Hehlen, M. P.},
  year = 2010,
  month = mar,
  journal = {Journal of the Optical Society of America B},
  volume = {27},
  number = {3},
  pages = {611},
  issn = {0740-3224, 1520-8540},
  doi = {10.1364/JOSAB.27.000611},
  urldate = {2025-05-20},
  copyright = {https://doi.org/10.1364/OA\_License\_v1\#VOR},
  langid = {english}
}

@article{Sahoo2022a,
  title = {Lattice-Strain Induced Photophysical Properties of {{NaYF4}}: {{Yb3}}+, {{Tm3}}+ Upconverting Phosphors},
  shorttitle = {Lattice-Strain Induced Photophysical Properties of {{NaYF4}}},
  author = {Sahoo, Kedar and Ranjan, Sudhir and Kumar, Manoj},
  year = 2022,
  month = nov,
  journal = {Journal of Luminescence},
  volume = {251},
  pages = {119249},
  issn = {00222313},
  doi = {10.1016/j.jlumin.2022.119249},
  urldate = {2025-05-19},
  langid = {english}
}

@misc{Sheldrick1997,
  title = {{{SHELXL-97}}, {{Program}} for the {{Refinement}} of {{Crystal Structures}}},
  author = {Sheldrick, George Michael},
  year = 1997
}

@misc{Sheldrick2005,
  title = {{{CELL}}\_{{NOW}}},
  author = {Sheldrick, George Michael},
  year = 2005
}

@article{Sheldrick2007_SHELX,
  title = {A Short History of {{SHELX}}},
  author = {Sheldrick, George M.},
  year = 2007,
  month = dec,
  journal = {Acta Crystallographica Section A Foundations of Crystallography},
  volume = {64},
  number = {1},
  pages = {112--122},
  publisher = {International Union of Crystallography (IUCr)},
  issn = {0108-7673},
  doi = {10.1107/s0108767307043930}
}

@misc{Sheldrick2007_TWINABS,
  title = {{{TWINABS}}},
  author = {Sheldrick, George Michael},
  year = 2007
}

@article{Sheldrick2015,
  title = {{{SHELXT}}-- {{Integrated}} Space-Group and Crystal-Structure Determination},
  author = {Sheldrick, George M.},
  year = 2015,
  month = jan,
  journal = {Acta Crystallographica Section A Foundations and Advances},
  volume = {71},
  number = {1},
  pages = {3--8},
  publisher = {International Union of Crystallography (IUCr)},
  issn = {2053-2733},
  doi = {10.1107/s2053273314026370}
}

@article{Sheldrick2015a,
  title = {Crystal Structure Refinement with {{SHELXL}}},
  author = {Sheldrick, George M.},
  year = 2015,
  month = jan,
  journal = {Acta Crystallographica Section C Structural Chemistry},
  volume = {71},
  number = {1},
  pages = {3--8},
  publisher = {International Union of Crystallography (IUCr)},
  issn = {2053-2296},
  doi = {10.1107/s2053229614024218}
}

@article{Smith1987,
  title = {Critical Stability Constants, Enthalpies and Entropies for the Formation of Metal Complexes of Aminopolycarboxylic Acids and Carboxylic Acids},
  author = {Smith, Robert M. and Martell, Arthur E.},
  year = 1987,
  month = jun,
  journal = {Science of The Total Environment},
  volume = {64},
  number = {1-2},
  pages = {125--147},
  issn = {00489697},
  doi = {10.1016/0048-9697(87)90127-6},
  urldate = {2025-11-10},
  copyright = {https://www.elsevier.com/tdm/userlicense/1.0/},
  langid = {english}
}

@article{Som2021,
  title = {Synthesis and Design of {{NaYF}}{\textsubscript{4}} Microprisms via a Microwave-assisted Approach for Highly Sensitive Optical Thermometry Applications},
  author = {Som, Sudipta and Lu, Chung-Hsin and Yang, Che-Yuan and Das, Subrata},
  year = 2021,
  month = oct,
  journal = {Journal of the American Ceramic Society},
  volume = {104},
  number = {10},
  pages = {5168--5181},
  issn = {0002-7820, 1551-2916},
  doi = {10.1111/jace.17783},
  urldate = {2025-05-19},
  langid = {english}
}

@article{Thoma1963,
  title = {Phase Equilibria in the System Sodium Fluoride-Yttrium Fluoride},
  author = {Thoma, R. E. and Hebert, G. M. and Insley, H. and Weaver, C. F.},
  year = 1963,
  month = oct,
  journal = {Inorganic Chemistry},
  volume = {2},
  number = {5},
  pages = {1005--1012},
  publisher = {American Chemical Society (ACS)},
  doi = {10.1021/ic50009a030}
}

@article{Waasmaier1995,
  title = {New Analytical Scattering-Factor Functions for Free Atoms and Ions},
  author = {Waasmaier, D. and Kirfel, A.},
  year = 1995,
  month = may,
  journal = {Acta Crystallographica Section A Foundations of Crystallography},
  volume = {51},
  number = {3},
  pages = {416--431},
  publisher = {International Union of Crystallography (IUCr)},
  issn = {0108-7673},
  doi = {10.1107/s0108767394013292}
}

@article{Wang2006,
  title = {Selected Synthesis of Cubic and Hexagonal {{NaYF4}} Crystals via a Complex-Assisted Hydrothermal Route},
  author = {Wang, Zhijun and Tao, Feng and Yao, Lianzeng and Cai, Weili and Li, Xiaoguang},
  year = 2006,
  month = apr,
  journal = {Journal of Crystal Growth},
  volume = {290},
  number = {1},
  pages = {296--300},
  publisher = {Elsevier BV},
  issn = {0022-0248},
  doi = {10.1016/j.jcrysgro.2006.01.012}
}

@article{Wang2013,
  title = {Synthesis of {{NaYF4}} Microcrystals with Different Morphologies and Enhanced Up-Conversion Luminescence Properties},
  author = {Wang, Yan and Gai, Shili and Niu, Na and He, Fei and Yang, Piaoping},
  year = 2013,
  journal = {Physical Chemistry Chemical Physics},
  volume = {15},
  number = {39},
  pages = {16795},
  issn = {1463-9076, 1463-9084},
  doi = {10.1039/c3cp52813h},
  urldate = {2025-05-15},
  langid = {english}
}

@article{Wang2015a,
  title = {Controlled Synthesis and Luminance Properties of Lanthanide-Doped {$\beta$}-{{NaYF4}} Microcrystals},
  author = {Wang, Junmei and Li, Kexun and Zhao, Yanbao and Zhu, Zhenping},
  year = 2015,
  month = apr,
  journal = {Journal of Rare Earths},
  volume = {33},
  number = {4},
  pages = {339--345},
  issn = {10020721},
  doi = {10.1016/S1002-0721(14)60423-3},
  urldate = {2025-05-19},
  copyright = {https://www.elsevier.com/tdm/userlicense/1.0/},
  langid = {english}
}

@article{Wang2017,
  title = {White-{{Light Whispering-Gallery-Mode Lasing}} from {{Lanthanide-Doped Upconversion NaYF4 Hexagonal Microrods}}},
  author = {Wang, Ting and Yu, Huan and Siu, Chun Kit and Qiu, Jianbei and Xu, Xuhui and Yu, Siu Fung},
  year = 2017,
  month = may,
  journal = {ACS Photonics},
  volume = {4},
  number = {6},
  pages = {1539--1543},
  publisher = {American Chemical Society (ACS)},
  issn = {2330-4022},
  doi = {10.1021/acsphotonics.7b00301}
}

@article{Winstone2022,
  title = {Optical {{Trapping}} of {{High-Aspect-Ratio NaYF Hexagonal Prisms}} for {{kHz-MHz Gravitational Wave Detectors}}},
  author = {Winstone, George and Wang, Zhiyuan and Klomp, Shelby and Felsted, Greg R. and Laeuger, Andrew and Gupta, Chaman and Grass, Daniel and Aggarwal, Nancy and Sprague, Jacob and Pauzauskie, Peter J. and Larson, Shane L. and Kalogera, Vicky and Geraci, Andrew A.},
  year = 2022,
  month = jul,
  journal = {Physical Review Letters},
  volume = {129},
  number = {5},
  pages = {053604},
  publisher = {American Physical Society (APS)},
  doi = {10.1103/physrevlett.129.053604}
}

@article{Wurth2018,
  title = {Quantum {{Yields}}, {{Surface Quenching}}, and {{Passivation Efficiency}} for {{Ultrasmall Core}}/{{Shell Upconverting Nanoparticles}}},
  author = {W{\"u}rth, Christian and Fischer, Stefan and Grauel, Bettina and Alivisatos, A. Paul and {Resch-Genger}, Ute},
  year = 2018,
  month = apr,
  journal = {Journal of the American Chemical Society},
  volume = {140},
  number = {14},
  pages = {4922--4928},
  issn = {0002-7863, 1520-5126},
  doi = {10.1021/jacs.8b01458},
  urldate = {2025-09-08},
  langid = {english}
}

@article{Yang2012a,
  title = {One-{{Step Hydrothermal Synthesis}} of {{Carboxyl}}-{{Functionalized Upconversion Phosphors}} for {{Bioapplications}}},
  author = {Yang, Jianping and Shen, Dengke and Li, Xiaomin and Li, Wei and Fang, Yin and Wei, Yong and Yao, Chi and Tu, Bo and Zhang, Fan and Zhao, Dongyuan},
  year = 2012,
  month = oct,
  journal = {Chemistry -- A European Journal},
  volume = {18},
  number = {43},
  pages = {13642--13650},
  issn = {0947-6539, 1521-3765},
  doi = {10.1002/chem.201202336},
  urldate = {2025-05-15},
  copyright = {http://onlinelibrary.wiley.com/termsAndConditions\#vor},
  langid = {english}
}

@article{Yi2007,
  title = {Water-{{Soluble NaYF}}{\textsubscript{4}} :{{Yb}},{{Er}}({{Tm}})/{{NaYF}}{\textsubscript{4}} /{{Polymer Core}}/{{Shell}}/{{Shell Nanoparticles}} with {{Significant Enhancement}} of {{Upconversion Fluorescence}}},
  shorttitle = {Water-{{Soluble NaYF}}{\textsubscript{4}}},
  author = {Yi, Guang-Shun and Chow, Gan-Moog},
  year = 2007,
  month = feb,
  journal = {Chemistry of Materials},
  volume = {19},
  number = {3},
  pages = {341--343},
  issn = {0897-4756, 1520-5002},
  doi = {10.1021/cm062447y},
  urldate = {2025-05-20},
  langid = {english}
}

@article{Zhang2022,
  title = {Luminescence Thermometry with Rare Earth Doped Nanoparticles: {{Status}} and Challenges},
  author = {Zhang, Baobao and Guo, Xiaojun and Zhang, Zhenglong and Fu, Zhengkun and Zheng, Hairong},
  year = 2022,
  month = oct,
  journal = {Journal of Luminescence},
  volume = {250},
  pages = {119110},
  publisher = {Elsevier BV},
  doi = {10.1016/j.jlumin.2022.119110}
}

@article{Zheng2022,
  title = {Rare-{{Earth Doping}} in {{Nanostructured Inorganic Materials}}},
  author = {Zheng, Bingzhu and Fan, Jingyue and Chen, Bing and Qin, Xian and Wang, Juan and Wang, Feng and Deng, Renren and Liu, Xiaogang},
  year = 2022,
  month = jan,
  journal = {Chemical Reviews},
  volume = {122},
  number = {6},
  pages = {5519--5603},
  publisher = {American Chemical Society (ACS)},
  issn = {1520-6890},
  doi = {10.1021/acs.chemrev.1c00644}
}

@article{Zhu2019,
  title = {Recent {{Progress}} of {{Rare-Earth Doped Upconversion Nanoparticles}}: {{Synthesis}}, {{Optimization}}, and {{Applications}}},
  author = {Zhu, Xiaohui and Zhang, Jing and Liu, Jinliang and Zhang, Yong},
  year = 2019,
  month = sep,
  journal = {Advanced Science},
  volume = {6},
  number = {22},
  pages = {1901358},
  publisher = {Wiley},
  doi = {10.1002/advs.201901358}
}

\end{document}